\documentstyle[aps,epsfig,pre,aps,epsf]{revtex}

\begin{document}
\draft
\input{psfig}
\title{Structure and rheology of binary mixtures in shear flow}
\author{F. Corberi}
\address{Istituto Nazionale per la Fisica della Materia, Unit\`a di Salerno
and Dipartimento di Fisica, Universit\'a di Salerno, 84081 Baronissi (Salerno),
Italy.}

\author{G. Gonnella and A. Lamura}
\address{Istituto Nazionale per la  Fisica della Materia, Unit\`a di Bari
{\rm and} Dipartimento di Fisica, Universit\`a di Bari, {\rm and}
Istituto Nazionale di Fisica Nucleare, Sezione di Bari, via Amendola
173, 70126 Bari, Italy.}
\date{\today}
\maketitle
\begin{abstract}
Results are presented for the phase separation process
of a binary mixture subject to an uniform shear flow quenched 
from a disordered to a homogeneous ordered
phase. The kinetics of the process is described in the context of   
the time-dependent Ginzburg-Landau equation with an external
velocity term. The large-$n$ approximation is used to study the evolution
of the model in the presence of a stationary flow and in the case
of an oscillating shear.

For stationary flow
we show that the structure factor obeys a generalized
dynamical scaling. The domains grow with different typical lengthscales
$R_x$ and  $R_\perp$ respectively 
in the flow direction and perpendicularly to it. In the 
scaling regime $R_\perp \sim t^{\alpha_\perp}$ and $R_x \sim \gamma  
t^{\alpha_x}$
(with logarithmic corrections),
$\gamma $ being the shear rate, with $\alpha _x=5/4$ and
$\alpha _\perp =1/4$. 
The excess viscosity $\Delta \eta$ after reaching a maximum 
relaxes to zero as $\gamma ^{-2}t^{-3/2}$. $\Delta \eta$ and other
observables exhibit log-time periodic oscillations which can be
interpreted as due to a growth mechanism where 
stretching  and break-up of domains cyclically occur.

In the case of an oscillating shear a cross-over phenomenon is observed: 
Initially the evolution is characterized by the
same growth exponents as for a stationary flow.
For longer times the phase separating structure cannot align with 
the oscillating 
drift and a different regime is entered with an isotropic growth  
and the same exponents of the case without shear. 

\end{abstract}

\section{Introduction}

The kinetics of phase separation of a  disordered system quenched
into a multiphase coexistence region has been extensively studied
in the last years \cite{Bin}. The main features
of the process are well understood:
After an early stage during which ordered 
domains of the equilibrium phases are formed
the segregation proceeds in the late stage 
by coarsening of ordered regions
according to the power law growth  
$R(t) \sim t^{\alpha}$ for the average domains size.
In binary liquids,  
the existence of several regimes characterized
by different exponents $\alpha$, due to the presence 
of various growth mechanisms, is well established \cite{B94}.
In these regimes the pair correlation function $C(r,t)$ verifies 
a dynamical scaling
law according to which it can be written as $C(r,t) \simeq f(r/R)$, 
where $f(x)$ is a scaling function \cite{yeo}. 

>From the theoretical point of view the most relevant progresses have been
achieved in the framework of the continuous approach based on the 
Cahn-Hilliard equation with a Ginzburg-Landau free energy functional,
the time dependent Ginzburg-Landau (TDGL) model.
Within this approach, which neglects hydrodynamics, 
the properties of the phase-separation kinetics
can be efficiently studied by means of numerical simulations or analytically
in the context of approximate theories, among which the so-called large-$n$
limit (one-loop approximation).
For a vectorial system with an infinite number of components
$n$, indeed, the TDGL model is exactly soluble.
The one-loop approximation is known to provide a mean-field picture of the 
phase-separation process which captures the essence of the phenomenon
at a semi-quantitative level \cite{crz}.
 
In this paper we study the process of phase separation in a binary
mixture subject to an uniform shear flow.  
When shear is applied to the system the time evolution is 
substantially different from that of ordinary spinodal decomposition.
We consider both a stationary flow
and an oscillating shear.
 
A stationary flow induces 
strong deformations of the domains formed
after the quench \cite{On97,OND,Ro}, which become anisotropic and 
stretched along the flow direction. 
Consequently the growth rate along the flow is larger than in the other
directions.
In some experiments a power law increase of the typical size of the
domains is observed and a value 
$\Delta \alpha  = \alpha_x  - \alpha_\perp$ in the range $0.8\div 1$ 
for the difference between the exponents in 
the flow and in the shear directions is reported \cite{LLG,Bey}.
Two dimensional molecular dynamic simulations 
find a slightly smaller value  
\cite{PT}.
In other experimental realizations, when the shear is strong enough, stringlike
domains have been observed to extend macroscopically
in the direction of the flow \cite{Has} preventing complete phase
separation. 
In general, the scaling behavior of sheared systems is not clearly understood
and the very existence of a scaling regime in different experimental systems
is questionable.  

In a previous paper \cite{noi} we have shown that the numerical solution
of the one-loop 
approximation to the TDGL model for
phase separation under shear exhibits a generalized 
scaling symmetry characterized
by  $\Delta \alpha  = 1$. 
In the scaling regime the structure factor and other observables
exhibit the interesting feature of an oscillatory pattern
which can be related to a mechanism of storing and dissipation
of elastic energy where domains are stretched and broken cyclically. 
This new effect has been shown to persist up to the longest
available time of our computation and represents the hallmark of
a complex dynamical pattern induced by the presence of the shear.
In a recent paper Rapapa and Bray \cite{rapa}, by solving 
asymptotically the one-loop equations, confirmed analytically 
the existence of a (multi)-scaling symmetry; in the long time
limit, however, they do not recover the cyclical pattern described
insofar and they infer 'that the observed oscillations are slowly-decaying 
preasymptotic transients'. Since their solution is obtained in the infinite
time limit, then, a reference theory for the description of this remarkable
phenomenon is lacking.

Given that the one-loop approximation is a mean-field solution in spirit the
natural question of its accuracy for the description of the original
model arises. 
A numerical analysis of the exact TDGL model
has been performed recently in \cite{noi2} where it is shown that
the global picture of the one-loop approximation is adequate.
In particular the oscillatory pattern is recovered.
The existence of a scaling symmetry and the determination of the
related exponents, however, has not been clearly established numerically
mainly due to finite size effects limitations.
The actual value of the growth 
exponents can be inferred by scaling \cite{rapa} or renormalization
group \cite{noi2} arguments to be $\alpha _\perp=1/3$, as in the case without
shear (we stress the fact that hydrodynamic effects are neglected in this 
model), and $\alpha _x=4/3$. 

The shear also induces a peculiar rheological behavior.
The break-up of the stretched domains liberates an energy which gives rise
to an increase $\Delta \eta$ of the viscosity  
\cite{Onu,KSH}. 
Experiments and simulations show that the
excess viscosity $\Delta \eta$ reaches a maximum at $t=t_m$ 
and then relaxes to smaller values.
The maximum of the excess viscosity is expected to occur at a fixed 
$\gamma t$  and to scale as  
$\Delta \eta (t_m) \sim \gamma^{-\nu}$ \cite{OND,LLG}.
Simple scaling arguments 
predict $\nu=2/3$ \cite{OND}, but different values 
have been reported \cite{LLG}. All these features are adequately 
described by the TDGL already at the one-loop level.

In this paper we present a complete scenario of the behavior of the TDGL model
for phase separation in a shear flow in the framework of the large-$n$
approximation. 
The behavior of the system is studied along the whole time
history, from the instant of the quench onward, both in the presence of a 
steady flow and in the case of an oscillating shear where interesting
effects are undercovered. Results are presented for two and three dimensional
systems.

This paper is schematically divided as follows:
In Sec.2 we specify the model and introduce the one-loop approximation that
will be studied thoroughly in the following Sections. Section 3 is devoted to 
the analysis of the behavior of the model subjected to a steady flow.
In Sec.4 the dynamics in the presence of an oscillatory shear is considered.
In Sec.5 we present a discussion of the results, debate some open problems  
and draw our conclusions.

\section{The model}

The binary mixture is described at equilibrium by a Ginzburg-Landau 
free-energy 
\begin{equation}
{\cal F}\{\varphi\} = \int d^d r 
\{\frac{a}{2} \varphi^2 + \frac{b}{4} \varphi^4 
+ \frac{\kappa}{2} \mid \nabla \varphi \mid^2 \}
\label{eqn1}
\end{equation}
where $\varphi$ is the order parameter which represents the concentration
difference between the two components. The values of  
$b,\kappa $ are positive for any temperature $T$ of the fluid.
The parameter $a$ separates stable states of the blend
with $a>a_c(T)$ ($a_c(T)\leq 0$), 
from the thermodynamically unstable states with $a<a_c(T)$
where the system phase separates.
The time evolution of the order parameter is given by the 
convection-diffusion equation
\begin{equation}
\frac {\partial \varphi} {\partial t} + \vec \nabla \cdot (\varphi \vec v) =
\Gamma \nabla^2  \frac {\delta {\cal F}}{\delta \varphi} + \eta
\label{eqn2}
\end{equation}
where the 
gaussian stochastic field $\eta$, with expectations
\begin{eqnarray}
\langle \eta (\vec r, t) \rangle &=& 0 \nonumber \\
\langle \eta(\vec r, t) \eta(\vec r', t')\rangle &=& 
                               -2 T \Gamma \nabla^2 \delta(\vec r - 
                               \vec r') \delta(t-t')
\label{eqn3}
\end{eqnarray}
describes thermal fluctuations \cite{On97}.
In Eq.~(\ref{eqn2}) $\Gamma$ is a transport coefficient 
and the symbol $\langle ...\rangle$ indicates the 
ensemble average.
The external velocity field here considered is of the form 
\begin{equation}
\vec v = \gamma y \vec e_x
\label{eqn4}
\end{equation}
where $\gamma$ is the spatially homogeneous shear rate \cite{On97},
that may however depend on time,
and $\vec e_x$ is a unit vector in the flow direction.
In the following we will consider a quench from an uncorrelated
isotropic high temperature initial condition
at the critical
composition, i.e. with $\langle \varphi (\vec r,0)
\rangle =0$ and $<\varphi(\vec r, 0) \varphi (\vec r',0)>=
\Delta \delta (\vec r-\vec r')$. 
The main observable for the description of the phase-separation kinetics
is the structure factor
\begin{equation}
C(\vec k,t) = <\varphi(\vec k, t) \varphi (-\vec k,t)>
\label{eqn5}
\end{equation}
where $\varphi(\vec k,t)$ is the Fourier transform of the
field $\varphi(\vec r, t)$ solution of  Eq. (\ref{eqn2}).
In the high temperature initial state we consider
one has $C(\vec k,0)=\Delta$.

The cubic term in the derivative $\delta {\cal F}/\delta \varphi$
prevents an exact solution of the Eq.~(\ref{eqn2}), 
as in the case without shear \cite{B94}.
However a soluble model is recovered in the one-loop approximation
which amounts to the factorization of the cubic term of Eq.~(\ref{eqn2}) 
as
\begin{equation}
\varphi^3 \to 
\langle \varphi^2\rangle \varphi
\label{oneloop}
\end{equation}
It is possible to show
\cite{Ma} that the substitution~(\ref{oneloop}) becomes exact
in models with a vectorial order parameter when the number $n$ of its 
components becomes infinite.
Since $\langle \varphi^2\rangle = S(t)$ does not depend on space, 
due to translational invariance, the substitution~(\ref{oneloop})
formally linearizes the theory.
The large-$n$ limit  is a well developed approximation scheme 
in statistical mechanics which have been applied to different contexts
\cite{varie}: Its
validity and limitations are nowadays rather well understood \cite{CCZ}. 

In the large-$n$ approximation the dynamical equation for 
$C(\vec k,t)$ is:
\begin{equation}
\frac {\partial C(\vec k,t)} {\partial t} - 
\gamma k_x \frac {\partial C(\vec k,t)} {\partial k_y} =    
- k^2 [k^2 + S(t) -1]C(\vec k,t) + k^2 T
\label{eqn6}
\end{equation}
where the function $S(t)$ is self-consistently given by
\begin{equation}
S(t)  =  \int _{|\vec k|<q}  \frac {d\vec k}{(2\pi)^d}  C(\vec k,t)
\label{eqn7}
\end{equation}
and $q$ is a high momentum phenomenological cut-off.
Notice that in Eq.~(\ref{eqn6}) the parameters of the free-energy
(\ref{eqn1}) and the mobility $\Gamma$ have been eliminated by a redefinition
of the time, space and field scales.
The rheological properties of the mixture are described in terms of
the shear stress 
\begin{equation}
\sigma _{xy}(t)= -\int _{|\vec k |<q}
\frac {d\vec k}{(2\pi)^d} k_x k_y C(\vec k,t) 
\label{eqn5bb}
\end{equation}
and of the first and second normal stress differences 
\begin{equation}
\Delta N_1 =  \int _{|\vec k |<q} \frac {d\vec k}{(2\pi)^d} 
[k_y^2- k_x^2] C(\vec k,t).
\label{eqn8}
\end{equation}
and
\begin{equation}
\Delta N_2 =  \int _{|\vec k |<q} \frac {d\vec k}{(2\pi)^d} 
[k_z^2- k_y^2] C(\vec k,t).
\label{eqn8bis}
\end{equation}
For vectorial systems with $n>d$ ($d$ is the spatial dimensionality)
topological defects are not stable \cite{B94}.
For large $n$, therefore, domains of the equilibrium phases are, strictly 
speaking, absent. Nevertheless, since from the solution of the one-loop 
equations presented below it is possible to identify characteristic 
growing lengths $R_x(t)$ and $R_\perp (t)$ in the flow and in the other 
directions it is natural to interpret these quantities as the {\it trace}
of the domains size after the one-loop approximation procedure
has been performed. In the following we will always use the word
{\it domains} in this broad acception.

\section{Steady shear}

In this section we consider the case of a constant shear rate $\gamma$.
Eq.~(\ref{eqn6}) can be formally integrated, yielding
\begin{equation}
C(\vec k,t)=\Delta e^{-\int _0 ^t {\cal K}^2(u)[{\cal K}^2(u)+S(t-u)-1]du}+
2T\int _0 ^t {\cal K}^2(u)
e^{-\int _0 ^u {\cal K}^2(s)[{\cal K}^2(s)+S(t-s)-1]ds}du
\label{formal}
\end{equation}
where 
\begin{equation}
\vec {\cal K}(u)=\vec k +\gamma k_x u \vec e _y
\label{kstorto}
\end{equation}
and $\vec e_y$ is the unit vector in the shear direction normal to the flow.
For steady flow it is usual to define the excess viscosity as
\begin{equation}
\Delta \eta (t)=\frac {\sigma _{xy}(t)}{\gamma}
\label{5b}
\end{equation}

\subsection{Analytic solution in the short and long time limit}

The consistency condition~(\ref{eqn7}) cannot be worked out along the whole
time history of the system. For this reason in the following Sections
the model Equations will be solved numerically both in $d=2$ and $d=3$.
However the model can be solved in the short and long time limit.

\vspace {1 cm}
{\bf Short times}
\vspace {1 cm}

For short times the linearized theory developed originally 
by Cahn and Hilliard \cite{CH} for the situation with $\gamma =0$ 
can be extended to the present case. This amounts to neglecting the  
quartic term in the local part of the free energy~(\ref{eqn1}) since
in the initial high temperature state the order parameter is small.
With this approximation the solution of Eq.~(\ref{eqn2}) reads
\begin{equation}
C(\vec k,t)=\Delta e^{-\int _0 ^t {\cal K}^2(z)[{\cal K}^2(z)-1]dz}+
2T\int _0 ^t {\cal K}^2(z) e^{-\int _0 ^z {\cal K}^2(s)[{\cal K}^2(s)-1]ds}
dz
\label{CahnHill}
\end{equation}
This approach 
applies to the original
model and to the large-$n$ approximation as well because non-linear terms
are neglected. 
It is well known that the linear theory describes 
the very initial transient of the phase-separation process, when domains
are still forming. In this time domain 
the behaviour of the system in the presence of the flow
is more interesting than in the simple case of an immobile fluid.
A plot of the structure factor~(\ref{CahnHill}) 
is presented for a two-dimensional system in Fig.~(\ref{fig_a}) for the case
$\gamma=1$ and $T=0$.
Initially, when domains are forming but the shear flow has not yet
produced sensible effects, the structure factor
evolves assuming the typical structure of a circular volcano, similarly to 
what happens in the case without shear. At  $\gamma t \simeq 0.5$ the 
anisotropy induced by the shear produces a 
deformation in the profile of the edge of the volcano from a ring-like
geometry into an ellipse, whose major axis forms with the positive
direction of the $k_{y}$ axis an angle of approximatively $45^o$
(see Fig.~(\ref{fig_a})). At the
same time small dips start to develop in the edge at the ends of the axes
of the ellipse and four peaks can be clearly observed at $\gamma t \simeq 2$.
As time goes by, the angle formed by the major axis of the ellipse with the
$k_y$-direction decreases and the dips in the profile of $C(\vec k,t)$ along 
the major axis develop until $C(\vec k,t)$ almost consists of
 two separated foils, at $\gamma t \simeq 4$, when a couple of
peaks prevails. 
The same initial pattern is also observed \cite{noi2}
by numerically solving the full model equation~(\ref{eqn2}).
At later times, however, the presence of the non-linear terms becomes 
fundamental and the linear theory breaks down, as in the case
without shear. It is important to stress the fact that the presence of four
peaks in the structure factor is exhibited already at the linear theory level
of approximation. We will see in the following sections that the very 
existence of a multiply peaked $C(\vec k,t)$ produces a rich dynamical
pattern originating an oscillatory phenomenon.

\vspace {1 cm}
{\bf Long times}
\vspace {1 cm}

The self-consistency condition~(\ref{eqn7}) has been worked out explicitly
in the long-time domain in \cite{rapa}. It is found that the model
is interested by a multiscaling symmetry, as in the case without shear
\cite {note}, characterized by the growth of the characteristic
lengthscales as 
\begin{equation}
R_x\sim \gamma (\frac{t^5}{\ln t})^{\frac{1}{4}} 
\label{rx}
\end{equation}
and
\begin{equation}
R_\perp \sim (\frac{t}{\ln t})^{\frac{1}{4}} 
\label{rperp}
\end{equation}
in the directions of the flow and perpendicular to it 
respectively. The excess viscosity and the normal stress differences 
behave as
\begin{equation}
\Delta \eta (t) \sim \gamma ^{-2} (\frac{\ln t}{t^3})^{\frac{1}{2}}
\label{ex}
\end{equation}
\begin{equation}
\Delta N_1 \sim \Delta N_2 \sim (\frac {\ln t}{t})^{\frac{1}{2}}
\label{diff}
\end{equation}
The same behaviors (apart from logarithmic corrections) is obtained in
\cite{noi} by means of a scaling ansatz.

\subsection{Numerical solution}

We present in this Section the results of the numerical integration of
the large-$n$ equation~(\ref{eqn6}) which allows to 
follow the whole time history of the phase-separation process.
 We restrict ourselves to the case with
$T=0$. 
An Euler first order discretization scheme has been implemented in 
$d=2$ and $d=3$ on $d$-dimensional lattices with $201$ mesh points per each 
direction.
For long times the structure factor is strongly peaked around typical
wavevectors which move towards zero as time goes on (see Fig \ref{fig_b}).
Given that the support of $C(\vec k,t)$ also shrinks to zero
it is possible to greatly improve the quality of the numerical
computation by using a self-adaptive mesh algorithm that follows the 
evolution of the support of the structure factor.
We have solved Eq.~(\ref{eqn6}) for various values of the shear rate
$\gamma $ in the range $[10^{-4},10^{-2}]$. 
We found that the qualitative behavior is the same for all the
values of $\gamma $ considered. 
>From the knowledge of the structure factor we compute the
characteristic lengths $R(t)$ as
\begin{equation} 
R_x(t)= \frac{1}{\sqrt {\overline {k_x^2}}}
\end{equation} where 
\begin{equation} 
\overline {k_x^2} = 
\frac{\int d\vec k k_x^2 C(\vec k,t)}{\int d\vec k  C(\vec k,t)}
\end{equation}
and the same for the other directions.

\vspace{1 cm}
{\bf $d=2$}
\vspace{0.4 cm}

The behavior of $C(\vec k,t)$ is shown in Fig.~(\ref{fig_b}) for $\gamma = 
0.001$. Initially
the evolution of the structure factor 
is resemblant to the one observed in 
Fig.~(\ref{fig_a})
where the linear theory for $C(\vec k,t)$ was plotted.
Later on, however, the linear theory fails because the non-linearities
become relevant, and the long-time regime is entered.
This is characterized by the 
shrinking of the support of $C(\vec k,t)$ towards the
origin with different rates for the shear and the flow directions so that the
tilt angle, namely the direction along which $C$ is aligned, decreases in time.
The structure factor is divided into two separated foils which are
symmetric due to the property $C(\vec k,t)=C(-\vec k,t)$. In
each foil two distinct peaks can be observed  
located at $(k_{x_1},k_{y_1})$ and $(k_{x_2},k_{y_2})$ with 
$|k_{x_1}| \simeq 2|k_{x_2}|$ and $|k_{y_1}| \simeq 2|k_{y_2}|$.
Their heights change in time. 
The first peak to prevail is that located at $(k_{x_1},k_{y_1})$, 
while the other peak dominates later. 
As time elapses the two peaks are observed to prevail alternatively.
This oscillatory behavior continues 
up to the longest times of our computations. 

In Fig.~(\ref{fig_c}) the quantities $(\gamma \ln t)^{1/4}R_x(t)$ and 
$(\gamma \ln t)^{1/4}R_y(t)$ are plotted against the strain $\gamma t$.
According to Eqs.(\ref{rx},\ref{rperp}) for long times these quantities
should collapse, for different values of the shear, on two power-law
mastercurves with exponents $5/4$ and $1/4$, respectively.
Here we observe that the collapse is indeed good, but the predicted
power-law behavior is modulated by an oscillatory pattern.
These oscillations are observed to be periodic on a log-time axis
and persist up to the limit of the computational time.

We now consider the rheological behavior of the mixture by plotting in 
Fig.~(\ref{fig_d})
the quantity $(\gamma /\ln t)^{1/2}\Delta \eta (t)$ against the strain.
This quantity reaches a maximum at $\gamma t \simeq 3.5$ and then decrease
as also found in experiments \cite{LLG}.
For long times Eq.~(\ref{ex}) would predict a data collapse for different
$\gamma $ on a single power-law master-curve with exponent $-3/2$.
Here the situation is similar to the previous figure, in that the predicted
behavior is modulated by log-time periodic oscillations.
On the bases of simple scaling arguments the maximum of the excess
viscosity $\Delta \eta (t_m)$ is expected to occur at a fixed 
$\gamma t$ and to scale as $\Delta \eta (t_m)\sim \gamma ^{-\nu}$,
with $\nu =2/3$ \cite{OND,LLG}.
These arguments do not directly apply to the one-loop
approximation since, due to the mean field nature, the exponents
are different. The asymptotic solution~(\ref{ex}) is not adequate to
this early stage effect. The $\gamma $ dependence of $\Delta \eta (t_m)$
is plotted in the inset of Fig.~\ref{fig_d} showing that a power-law behaviour
with $\nu \simeq 0.6$ is obeyed, in partial agreement with the aforementioned
scaling arguments.

In Fig.~(\ref{fig_e}) 
we report the numerical results for the first normal stress 
by plotting $(\gamma \ln t)^{-1/2}\Delta N_{1}$ against $\gamma t$ with
$\gamma =0.01$.
We find that $\Delta N_{1} (t)$ scales asymptotically as predicted
by  Eq.~(\ref{diff}) again modulated
by an oscillatory pattern.

The periodic oscillations observed in all the physical observables are due
to the competition between the different peaks of  $C(\vec k,t)$.
Let's refer to the behavior of the excess viscosity to understand how this 
competition affects the rheological quantities, using the features of the 
structure factor to obtain information about the domains evolution under 
the action of shear. 
$\Delta \eta$ reaches its first maximum when the shape of $C(\vec k,t)$
is such that the peak
located at $(k_{x_1},k_{y_1})$
prevails and the difference between
the height of the two peaks is maximal. 
At this time the domains are elongated by the flow and there is a 
prevalence of thin domains in the system. 
As these string-like domains are stretched further,
they eventually break up into two or more domains, dissipating the stored 
energy. This has two effects: the excess viscosity decreases
and, on the other hand,
 the thick domains, which have not yet been broken, prevail. 
In this situation 
the other peak of $C(\vec k,t)$ (which is located at $(k_{x_2},k_{y_2})$
and represents the smaller features) 
grows faster until it prevails and
$\Delta \eta$ reaches a minimum. This behavior is reproduced 
with a characteristic frequency in log-time.
Recently, a similar behavior has been observed
in the numerical simulation of the full model 
Equation~(\ref{eqn2}) \cite{noi2}. 

\vspace{1 cm}
{\bf $d=3$}
\vspace{0.4 cm}

In this section we report the result of the numerical solution of 
Eq.~(\ref{eqn6}) in $d=3$.
In Fig.~(\ref{fig_f}) the time
evolution of the structure factor in the special planes $k_{x}=0$, $k_{y}=0$ 
and $k_{z}=0$ is shown for $\gamma=0.001$. In the plane $k_{z}=0$
$C(\vec k,t)$ behaves analogously to the previously
discussed two dimensional case. 
The structure factor on the plane $k_{y}=0$ 
gives information relative to the observation of the system along the shear 
direction: No velocity gradient is present in the plane perpendicular 
to this orientation, but there are different
velocities in the $x$ and $z$ directions.
This allows to explain the observed
behavior which is rather different from the one observed at  $k_{z}=0$. 
The structure factor develops initially a circular
volcano, as without shear. The edge of the volcano is progressively
deformed by the shear into an ellipse with the major axis along the $k_{z}$
direction. The dips in the edge of the volcano at values of $k_{x}
\simeq 0$ develop with time so that at $\gamma t \simeq 1$, $C(\vec k,t)$
is made of two foils but these are not completely separated. 
During the time evolution the
axes of the ellipse shrink; the decrease is faster along the $k_{x}$
direction. The two foils are never completely separated  and the 
angle formed with the $k_{z}$ direction is zero, as observed 
in experiments \cite{Bey}. At $\gamma t
\simeq 5$ two well-formed peaks start to develop and grow 
on each foil of $C(\vec k,t)$. These four peaks have the same height and
their relative heights do not change in
time, as it can be seen at $\gamma t=20$ in the picture, 
differently from the situation on the $k_z=0$ plane. 

In the $k_{x}=0$
plane  the shear has no effect at all and the structure factor 
remains circular during its evolution.

The computed behavior of $R_x(t)$ and $R_y(t)$ is similar 
to that of the two-dimensional case.
We also find $R_z(t)\sim R_y(t)$, as expected.

We report in Fig.~(\ref{fig_g}) the plots
of $(\gamma /\ln t)^{1/2}\Delta \eta (t)$, 
 $(\gamma \ln t)^{-1/2} \Delta N_1 (t)$ and 
 $- (\gamma \ln t)^{-1/2} \Delta N_2 (t)$
as functions of $\gamma t$. 
It appears that the rheological quantities still have amplitudes which are
modulated by log-time oscillations which are in phase among them.
The origin of such oscillations has
to be found again in the oscillations of the peaks of the structure factor in 
the plane $k_{z}=0$. 
Since the support of $C(\vec k,t)$ shrinks towards the origin
faster in the $k_{z}$ than in the $k_y$ direction 
the second normal stress difference
$\Delta N_2(t)$ is negative. This is in
accordance with general experimental experience \cite{Barnes}.

\section{Oscillating shear}

In this Section we consider 
the case of 
a time-dependent shear rate with
\begin{equation}
\gamma (t) =\gamma_{o} \cos \omega t
\label{gam}
\end{equation}
This situation is of great experimental relevance expecially for probing the
viscoelastic properties of the phase separating binary mixture. 

We solved the Eq.~(\ref{eqn6})
numerically in $d=2$ using the same 
numerical scheme as in the case of steady shear,
for different values of $\gamma _0$ and 
$\omega $.
We will describe below the case 
$\gamma _0 =10^{-3}$, $\tau =2 \pi/\omega=6 \times 10^3$.
The time evolution of the structure factor in the first cycle of $\gamma (t)$
is shown in Fig.~(\ref{fig_h-1}). 
The dynamical pattern is analogous to the one with $\gamma = const.$ for
times $t< \tau/4 $ as it can be seen at
$\gamma_{o} t=1.5$. 
Then the time dependent velocity field modifies the behaviour of the blend 
with respect to the case of a steady flow. In particular,
at the end of the first oscillation, the four peaks of $C(\vec k,t)$  
are located at comparable distances from the origin of the $k$-space
differently from what observed in Fig.~(\ref{fig_b}) at $\gamma t=6$.
The two highest maxima at $\gamma _0 t=6$ in Fig.~(\ref{fig_h-1})
are characterized by $|k_y| \gg |k_x|$. 
During the later time evolution these peaks  
grow and move towards the origin. The position in
the $k$-plane of the other peaks 
rotates back and forth cyclically along an approximatively circular
path. The radius of this trajectory shrinks towards the origin 
at a rate comparable with that of the position of the other peaks. 
In the asymptotic regime the four peaks
have approximatively the same height 
and the cyclical rotation of the peak position persists.
This is shown in Fig.~(\ref{fig_h}) where the configurations 
of the structure factor are shown at
each quarter of oscillation of the shear rate in the 
asymptotic stage. 

In Fig.~(\ref{fig_i}) 
the evolution of the characteristic lengths $R_x(t)$, $R_y(t)$
is plotted against $\gamma _0 t$. 
We also plot, in the inset, the time average of these quantities over a period
$\tau $, in order to smoothen out the superimposed oscillations. 
Here we observe, for times $t<\tau $, growth laws analogous to
the steady shear case, namely $R_x(t)\sim t^{5/4}$ and $R_y(t)\sim t^{1/4}$. 
The growth exponent of $R_{x}$ changes smoothly, from $t\simeq \tau $ 
onward, from $5/4$ to the asymptotic value $1/4$
which is reached at $\gamma_{o} t \sim 80$ when all the four peaks of the 
structure factor have the same height.
The gradual crossover of $\alpha _x$ from $5/4$ to $1/4$ 
can be better observed for larger values of $\tau $, 
since the regime with $\alpha_{x}=5/4$ persists for a longer time.
This is shown in Fig.~(\ref{fig_i+1}), 
where the evolution of $R_x(t)$, $R_y(t)$
is plotted against $\gamma _0 t$ for $\tau = 5 \times 10^5$.
For small $\tau $, instead, $R_{x}$ and $R_{y}$ grow with the same
exponent $1/4$ from the beginning.

These observations suggest the following physical interpretation:
for $t<\tau /2$, since $\gamma$ does not change sign,
the evolution of the blend is comparable to the case with a constant
shear rate. In particular, if $\tau $ is sufficiently large
to exceed the initial stage when domains are forming, the power
growth laws described in Sec.~3 for $R_x,R_\perp$ are observed with
$\alpha _x=5/4$ and $\alpha _{\perp}=1/4$.
On timescales much longer then $\tau $, however, 
the network of the larger domains cannot
be efficiently tilted along the flow orientation which changes
periodically its sign. This is confirmed by the behaviour of
the two peaks with $|k_y| \gg |k_x|$ whose position in the $k$-plane
moves toward the origin but does not cross the $k_x=0$ plane,
as it would be the case if the orientation of the domains
corresponding to these peaks were reversed.
In this situation the difference 
$\Delta \alpha =1$ between the exponents in the flow and shear direction 
cannot be sustained, because the larger domains are not
directed along the flow orientation at all times,  
and a growth law with the same exponent 1/4 in all the directions
is obeyed. It is interesting to notice that the other peaks, 
which represent smaller domains
formed by the break-up of the larger ones,
crosses the $k_y=0$ plane during their rotation every half period
of $\gamma$. This suggests that these features are tilted by the oscillating
shear and follow the flow orientation. 
Then we expect to observe in a real blend two type of domains which respond
differently to the oscillations of the flow: a network of large
and elongated structures which maintains the orientation imposed
during the first half period of $\gamma$ and a multitude of more isotropic
features, generated by the break-up of strained regions, which
oscillate following the flow. 
 
For studying rheological properties it is customary 
\cite{Krall} to introduce a complex viscosity
$\eta^{*} \equiv \eta ' - i \eta ''$ which is related to the 
shear stress by
\begin{equation}
\sigma_{xy} (t) =\gamma_{o} (\eta ' \cos \omega t+ \eta '' \sin \omega t)
\label{stress}
\end{equation}
when Eq.~ (\ref{gam}) holds. 
It is also useful to consider \cite{PD} another 
representation of the shear stress given by 
\begin{equation}
\sigma_{xy} (t) =C \sin(\omega t + \phi)
\label{stressbis}
\end{equation}
The connection between Eqs.~ (\ref{stress}) and  (\ref{stressbis}) is given by
\begin{equation}
C = \gamma_{o} \sqrt{\eta '^2 + \eta ''^2}
\label{ampl}
\end{equation}
and  
\begin{equation}
\tan \phi = \frac{\eta '}{\eta ''} .
\label{phase}
\end{equation}
By defining $\gamma^{*} (t)=
\gamma_{o} e^{i \omega t}$, we can write Eq. (\ref{stress}) as
\begin{equation}
\sigma_{xy} (t) =
Re \big [ \eta^{*} \gamma^{*} (t) \big ]
\end{equation}
In order to relate the real and
imaginary parts of the viscosity to physical quantities 
Eq.~(\ref{stress}) can be casted as
\begin{equation}
\sigma_{xy} (t) =\eta \gamma (t)+ G \int_{0}^{t} \gamma (t') dt'
\label{shstress}
\end{equation}
where $\eta=\eta '$, $G=\omega \eta ''$ and the  
identity $\sin \omega t = \omega \int_{0}^{t} \cos \omega t' dt'$ has
been used.

The coefficient $\eta$ in the r.h.s. of the Eq.~ (\ref{shstress})
multiplies the portion of the shear stress in phase with the shear rate
and represents the viscosity of a viscoelastic fluid. The integral
of the second term of the r.h.s. of the Eq.~ (\ref{shstress}) can be identified
with the shear strain present in the mixture at time $t$. The
coefficient $G$ is therefore the effective elastic shear modulus of the fluid.
Pure viscous behavior corresponds to $G=0$
($\phi=\pi/2$), pure elastic behavior to $\eta=0$ ($\phi=0$) \cite{PD}.

In order to compute $\eta$ and $G$ during the phase separation
we calculated by numerical integration
the shear stress using its general definition~(\ref{eqn5bb}).
By writing 
\begin{equation}
\sigma_{xy} (t) =A \cos \omega t+ B \sin \omega t
\label{sh}
\end{equation}
it follows that $\eta=A (t)/\gamma_{o}$ and $G=\omega B (t)/\gamma_{o}$.
In general, $A$ and $B$ depend on time. During a single shear
oscillation, however, we expect that the Eq.~ (\ref{sh}) holds as a good
approximation with constant values for $A$ and $B$. In this way $\sigma_{xy}
 (t)$ is expressed in terms of the first two coefficients in a Fourier
series expansion over the interval of scaled time of duration $2 \pi$.
The values we obtain may be referred to the time that locates the middle
of the interval. Thus we get
\begin{eqnarray}
\eta \left ( \left ( m-\frac{1}{2}\right ) \tau \right )
&=& \frac{1}{\gamma_{o} \pi}\int_{(m-1)\cdot 2 \pi}^{m \cdot 2 \pi}
\sigma_{xy} (t/\omega) \cos t \;dt \\ 
\label{visco}
G \left ( \left ( m-\frac{1}{2}\right ) \tau \right )
&=& \frac{\omega}{\gamma_{o} \pi}\int_{(m-1)\cdot 2 \pi}^{m \cdot 2 \pi}
\sigma_{xy} (t/\omega) \sin t \;dt 
\label{elastic}
\end{eqnarray}
where $m=1,2,..$.

In Fig.~(\ref{fig_l}) 
we report the plots of $\eta $ and $G$ against $\gamma_{o} t$.
The viscosity shows a crossover
between a power law decay with exponent $-3/2$ at short times and an
asymptotic behavior whose exponent is $-1/2$. 
This can  be explained observing that,
for the steady shear case,
the dynamic viscosity $\eta$ coincides with the excess viscosity 
which scales with the inverse of
the domains volume $V$. When an oscillatory shear is applied $V$ crosses over
from an initial power law increase $V\sim t^{3/2}$, similar to the one
for the case with steady shear (see Fig. (\ref{fig_c})),
 to a slower growth
$V\sim t^{1/2}$, as already discussed above for $R_x(t)$, 
producing a corresponding
crossover in $\eta$. 

>From the computed values of $\eta $ and $G$ we estimated the phase angle 
$\phi$, which, according to Eqs.~ (\ref{phase}) and (\ref{shstress}),
 is given by
$\displaystyle \phi=\arctan \Big (\frac{\eta}{G} \;\omega\Big)$. 
In Fig.~ (\ref{fig_m})
we report the time evolution of $\phi$ as a function of $\gamma_{o} t$.
It can be seen that $\phi$ decreases with time to reach an asymptotic value 
which is approximately 0.016. Accordingly, the system we are investigating
shows in the asymptotic stage a behavior which is essentially elastic.
The experimental data of \cite{Krall2} confirm this behavior.

\section{Summary and discussion}

In this paper we have studied the kinetics of a phase-separating
binary fluid, in the presence of a shear flow, by means of the
TDGL model. It is nowadays well established that
the corresponding model with $\gamma =0$ accurately describes the main 
features 
of the segregation process
in binary alloys, where hydrodynamics can be neglected.
In viscous fluids, such as polymeric blends, 
the validity of the present approach is limited to the early 
stage of spinodal decomposition; 
for longer times one should consider the full hydrodynamic 
description \cite{B94}.

When the shear is applied to a fluid the behavior of the system is
profoundly changed under many respects and a general agreement
on the predictivness of the proposed models is matter of general debate.
A discussion on possible effects of hydrodynamics is presented 
in \cite{yeo}.
Moreover, the numerical solution of the TDGL model with shear poses
serious problems due to discretization limitations and finite size
effects and, although some progresses have been recently achieved 
\cite{noi2},
a satisfactory description is nowadays not available.
In this scenario it is important to devise a simple analytical scheme
providing the fundamental tools for the comprehension of
the fluid dynamics. A natural choice in the field of growth kinetics
is the large-$n$ approximation, that has been thoroughly studied in the
case without shear, where it has proven to give a reliable description
of the segregation process, although at a {\it mean field} level.

In this paper the behavior of the TDGL model in the 
one-loop approximation is studied in detail,
and the whole time evolution of the blend is considered, from the
quenching instant onward; the cases of a
stationary flow and of an oscillating shear have been examined. 
In doing so we undercover a very rich dynamical 
pattern, where not only some experimental findings are reproduced, but
new predictions are allowed. 
After an early stage, which is accurately
described by the linear theory {\it \'a la} Cahn-Hilliard, the presence
of the velocity field produces an anisotropic power-law growth of the
characteristic lengths $R_x$, $R_\perp$ respectively in the flow direction and 
perpendicularly to it. 
The value of the exponent $\alpha_\perp =1/4$ in the directions perpendicular 
to the flow is the same as in models with vectorial conserved order parameter 
without shear; although the actual value of this exponent is not expected to be
accurate for real fluids (since, even without shear, the exponent 
obtained at the same level of approximation is known to corresponds to
the Lifshitz-Slyozov exponent $\alpha =1/3$ for scalar fields)  
a growth exponent $\alpha _\perp $
unaffected by the presence of shear has been obtained also by 
scaling \cite{rapa} and renormalization
group \cite{noi2} arguments applied to the 
full model equations and is also measured  in experiments 
\cite{LLG}.
Moreover a difference $\Delta \alpha =1$ between the flow and shear exponents
is also expected to be obtained by releasing the present approximation
\cite{rapa,noi2} 
and is observed in some experiments.
In the case of a stationary flow the anisotropic growth governed by these 
exponents is observed from the onset of the scaling regime onwards.
The power law behavior of any observable is decorated by log-time periodic
oscillations. 
These oscillations characterize the scaling
regime up to the longest simulated time but they are not
observed in the asymptotic solution presented in \cite{rapa}. 
Given that log-time periodicity appears to be a rather common feature
being observed, besides segregating fluids, during fracturing of
heterogeneous solids \cite{blume,sahimi} 
and in stock market indices \cite{sorn2} for instance, it would 
then be interesting to devise an analytical approach to enlighten
the origin of this new phenomenon at least in the present model.

In experiments with real fluid systems carried out by 
Laufer et al. \cite{Laufer} and, successively, by Mani et al. \cite{Mani}
and Migler et al. \cite{Migler} a {\it double overshoot}
in the time behavior of the viscosity and of the normal stress is
observed and an interpretation in terms of break-up and recombination of
the domains network is proposed. On the bases of our results it is
plausible that this {\it double overshoot} represents the first part of
a log-time periodic phenomenon which could be hopefully detected with
a suitable experimental setup. 
In the model we  have studied the oscillatory
behaviour is due to the competition between the different maxima of a
four-fold peaked structure factor. The presence of these maxima is
interpreted is Sec.~III B as due to the existence of different 
types of domains and the recurrent prevalence of each peak 
is suggested to be caused by the interplay between these kind of regions.
A structure factor with four maxima has also been observed in polymer mixtures
\cite{Migler}; however to our knowledge the connection between the alternative
dominance of the peaks of $C(\vec k,t)$ and the {\it overshoots} observed in
the viscosity and in the stresses has never been discussed insofar,
perhaps due to insufficient resolution, although an experimental
confirmation of this hypotheses would be desirable.

When an oscillating shear is present, the anisotropic regime
discussed so far for the steady shear case crosses over
to an isotropic growth when domains are fully developed. In this late stage,
from the analysis of the behavior of the structure factor, we conjecture
again the existence of two types of domains responding differently
to the flow oscillations: the network of elongated structures keeps the
orientation assumed during its formation in the early stage while
small features generated by scission of strained parts oscillate 
in phase with the flow.  
In this late stage the growth 
kinetics is regulated by the same exponents as without flow. We are not
aware of experiments reporting these features: It would be
interesting to devise an experimental set-up for testing
this prediction.

\acknowledgments
F.C. is grateful to M.Cirillo, R.Del Sole and M.Palummo for hospitality 
in the University of Rome.
F.C. acknowledges support by the TMR network contract ERBFMRXCT980183
and by PRA-1999 INFM and MURST(PRIN 97).

\newpage
\begin{center}
{\large CAPTIONS}
\end{center}
\begin{figure}
\caption{The evolution of the structure factor in the linear approximation
for $\gamma = 1$. The range of $k_x$ and $k_y$ varies from -2 to 2.
At $\gamma t=4$ the highest peaks are located at $(k_x,k_y) \simeq 
(0.0, \pm 0.75)$, the other two at $(k_x,k_y) \simeq 
(\mp 0.38, \pm 0.80)$.}
\label{fig_a}
\end{figure}
\begin{figure}
\caption{The evolution of the structure factor from the numerical solution
of Eq.~(\ref{eqn6}) in $d=2$ for $\gamma = 0.001$ and $T=0$. 
The range of $k_x$ and $k_y$ varies, at increasing times, as
$-0.6 \leq k_x, k_y \leq 0.6$ at $\gamma t=0.05, 1$;
$-0.15 \leq k_x \leq 0.15$ and $-0.6 \leq k_y \leq 0.6$ at $\gamma t=6, 8$;
$-0.075 \leq k_x \leq 0.075$ and $-0.6 \leq k_y \leq 0.6$ at $\gamma t=10$;
$-0.01875 \leq k_x \leq 0.01875$ and $-0.3 \leq k_y \leq 0.3$ 
at $\gamma t=45$.}
\label{fig_b}
\end{figure}
\begin{figure}
\caption{Data collapse (scaling plot) for the domains radii in $d=2$. 
The quantities $(\gamma \ln t)^{1/4}R_x(t)$ and 
$(\gamma \ln t)^{1/4}R_y(t)$ are plotted against the strain $\gamma t$
for different values of the
shear rate: $(\triangle) ~ \gamma=0.0001$, $(\circ) ~ \gamma=0.001$, 
$(\bullet) ~ \gamma=0.01$.
The two straight lines have slope 5/4 and 1/4.}
\label{fig_c}
\end{figure}
\begin{figure}
\caption{Data collapse (scaling plot) for the excess viscosity in $d=2$.
The quantity $(\gamma /\ln t)^{1/2}\Delta \eta (t)$ 
is plotted against the strain
$\gamma t$ for different values of the
shear rate: $(\triangle) ~ \gamma=0.0001$, $(\circ) ~ \gamma=0.001$, 
$(\star) ~ \gamma=0.005$, $(\bullet) ~ \gamma=0.01$.
The straight line has slope -3/2. The inset shows the maxima of $\Delta \eta$
as a function of $\gamma$. The slope of the straight line is 0.6.}
\label{fig_d}
\end{figure}
\begin{figure}
\caption{The first normal stress $\Delta N_1$ multiplied by
$(\gamma \ln t)^{-1/2}$ as a function of 
$\gamma t$. The value of $\gamma$ is 0.01. 
The slope of the straight line is -1/2.}
\label{fig_e}
\end{figure}
\begin{figure}
\caption{The structure factor from the numerical solution
of Eq.~(\ref{eqn6}) in $d=3$ at consecutive times for $\gamma = 0.001$
and $T=0$. From up to bottom the sections $k_{x}=0$, $k_{y}=0$ and $k_{z}=0$
of the structure factor are shown.
The range of $k_x$, $k_y$  and $k_z$ varies as $-0.6 \leq k_x, k_y,
k_z \leq 0.6$ at $\gamma t=1$; $-0.0375 \leq k_x \leq 0.0375$, 
$-0.3 \leq k_y \leq 0.3$ and $-0.3 \leq k_z \leq 0.3$ at $\gamma t=20$.}
\label{fig_f}
\end{figure}
\begin{figure}
\caption{$(\gamma /\ln t)^{1/2}\Delta \eta (t)$ ($\ast$), 
 $(\gamma \ln t)^{-1/2} \Delta N_1 (t)$ ($\circ$) and 
 $- (\gamma \ln t)^{-1/2} \Delta N_2 (t)$ ($\bullet$) as functions
of the strain
$\gamma t$.
The straight lines have slope $-3/2$ and $-1/2$.}
\label{fig_g}
\end{figure}
\begin{figure}
\caption{The structure factor at each quarter of the first $\gamma (t)$ 
oscillation. The range of $k_x$ and $k_y$ varies, at increasing times, as
$-0.6 \leq k_x, k_y \leq 0.6$ at $\gamma t=1.5$;
$-0.3 \leq k_x \leq 0.3$ and $-0.6 \leq k_y \leq 0.6$ at $\gamma t=3, 4.5$;
$-0.3 \leq k_x, k_y \leq 0.3$ at $\gamma t=6$.}
\label{fig_h-1}
\end{figure}
\begin{figure}
\caption{The structure factor at each quarter of a single oscillation in
the asymptotic stage. The range of $k_x$ and $k_y$ is
$-0.15 \leq k_x, k_y \leq 0.15$ at each time.}
\label{fig_h}
\end{figure}
\begin{figure}
\caption{The size of the domains in the $x$ (upper curve) and $y$ 
(lower curve)
directions are plotted against $\gamma _0 t$ for 
$\tau = 6 \times 10^3$. In the inset the same
quantities ($\ast$ and $\bullet$ for the $x$ and $y$ direction, respectively)
averaged over every single period of
oscillation are plotted at times $(m-1/2)\gamma_{o} \tau $, with
$m=1,2,..$, against $\gamma _0 t$.
The two straight lines have slope $1/4$.}
\label{fig_i}
\end{figure}
\begin{figure}
\caption{The size of the domains in the $x$ (upper curve) and $y$ 
(lower curve)
directions in the case  
$\tau = 5 \times 10^5$. 
The two straight lines have slope $5/4$ and $1/4$.}
\label{fig_i+1}
\end{figure}
\begin{figure}
\caption{The viscosity $\eta$ ($\ast$) and the elastic
shear modulus $G$ ($\bullet$) for $\tau = 6 \times 10^3$
at times $(m-1/2)\gamma_{o} \tau $ with
$m=1,2,..$.
The straight lines have slope $-3/2$ and $-1/2$.}
\label{fig_l}
\end{figure}
\begin{figure}
\caption{The time evolution of the phase angle $\phi$ 
for $\tau = 6 \times 10^3$
at times $(m-1/2)\gamma_{o} \tau $ with
$m=1,2,..$.}
\label{fig_m}
\end{figure}
\end{document}